\documentclass[journal]{IEEEtran}
\usepackage{graphicx}
\usepackage{subfigure}
\usepackage{latexsym}
\usepackage{amsmath}
\usepackage{multirow}
\usepackage{array}
\usepackage{algorithm}
\usepackage{algorithmic}
\begin{document}

\title{On Energy Efficiency and Delay Minimization in Reactive Protocols in Wireless Multi-hop Networks}

\author{N. Javaid$^{\ddag}$, U. Qasim$^{\pounds}$, Z. A. Khan$^{\$}$, M. A. Khan$^{\ddag}$, K. Latif$^{\ddag}$, A. Javaid$^{\natural}$\\\vspace{0.4cm}
        COMSATS Institute of IT, $^{\ddag}$Islamabad, $^{\natural}$Wah Cantt, Pakistan. \\
        $^{\$}$Faculty of Engineering, Dalhousie University, Halifax, Canada.\\
        $^{\pounds}$University of Alberta, Alberta, Canada.
     }

\maketitle

\begin{abstract}
In Wireless Multi-hop Networks (WMhNs), routing protocols with energy efficient and delay reduction techniques are needed to fulfill users' demands. In this paper, we present Linear Programming models (LP\_models) to assess and enhance reactive routing protocols. To practically examine constraints of respective LP\_models over reactive protocols, we select AODV, DSR and DYMO. It is deduced from analytical simulations of LP\_models in MATLAB that quick route repair reduces routing latency and optimizations of retransmission attempts results efficient energy utilization. To provide quick repair, we enhance AODV and DSR. To practically examine the efficiency of enhanced protocols in different scenarios of WMhNs, we conduct simulations using NS-2. From simulation results, enhanced DSR and AODV achieve efficient output by optimizing routing latencies and routing load in terms of retransmission attempts.
\end{abstract}


\section{Background}

In \cite{1}, authors evaluate AODV \cite{2}\cite{3} and DSR \cite{4}\cite{5} with respect to the varying number of Constant Bit Rate (CBR) resources. The authors in \cite{6}, evaluate performance of DSR and AODV with varying number of sources  ($10$ to $40$ sources with different pause times). Problem from a different perspective in \cite{7}, using simulation model with a dynamic network size and is practically examined for Destination-Sequence Distance Vector (DSDV) \cite{8}, AODV, DSR and Temporally-Ordered Routing Algorithm (TORA)\cite{9}.

In WMhNs, reactive protocols are responsible to find accurate routes and provide quick repair after detecting route breakages. This work is devoted to study routing capabilities of three reactive protocols named as AODV, DSR and DYnamic MANET On-demand (DYMO) \cite{10} in different network cases of WMhNs. The contribution of this work includes: \textit{(i)} construction of LP\_model for WMhNs requirements and analytical simulations of the models for selected protocols, \textit{(ii)} enhancements in AODV and DSR, \textit{(iii)} performance evaluation of the selected routing protocols with respect to framework of network constraints \textit{(iv)} analytical analysis of mobility, traffic rates and scalability properties of the selected routing protocols using NS-2.

\section{Reactive Protocols with their basic Operations}

The protocols use two basic operations; RD and RM. The total Energy Cost ($CE$) for reactive protocol, $rp$; $CE_{total}^{rp}$ \cite{11}:

\begin{equation}
CE_{total}^{rp}= CE_{RD}^{rp} + CE_{RM}^{rp}
\end{equation}
where, $CE_{RD}^{rp}$ and $CE_{RM}^{rp}$ represent energy cost for RD and RM processes, respectively.

\subsection{$CE_{RD}^{rp}$}

Expanding Ring Search (ERS)\cite{12}\cite{13} is used as optimization techniques in AODV, DSR and DYMO during RD. In ERS,  flooding is controlled by Time-To-Live (TTL) values to limit the broadcast. A source node $S_n$ may receive RREPs from the nodes that contain alternate (short) route for the desired destination $D_n$. $S_n$ establishes a path to $D_n$ which contains $9$ hops. These replies are only used in AODV and DSR and are known as gratuitous RREPs (\textit{grat. RREPs}). The destination RREPs are generated by the $D_n$ (destination RREPs are generated in all the three reactive protocols). So, control packet cost for RD; $CE_{RD}^{rp}$ can be calculated as \cite{13}:

\begin{equation}
CE_{RD}^{rp}= \sum_{i=1}^{M}CE_{k}(i)
\end{equation}

Let $M$ is number for maximum rings during $RD$. The generation of RREP(s) in AODV and DSR is also due to the valid routes in Routing Table $(RT)$ or in Route Cache $(RC)$, so, $M$ for DSR and AODV can be less than DYMO, because of absence of \textit{grat. RREPs} in DYMO. Let $d_{avg}$ is the average degree of nodes. The cost of any $k(i)$ can be calculated as $CE_{k}(i)$:

\begin{equation}
CE_{k}(i)= d_{avg} + d_{avg} \sum_{i=1}^{N_k} i
\end{equation}
where, $N_k$ represents the total number of nodes in ring $k_i$.

\subsection{$CE_{RM}^{rp}$}

In $RM$ process, different protocols pay different costs for link monitoring and also there are different costs for different supplementary maintenance strategies in case of link breakages. DYMO and AODV generate HELLO messages to check the connectivity of RN, while DSR gets the link level feedback from link layer.

In DYMO, link breakages in networks cause broadcasting of RERR messages. When the probability of unsuccessful local link repair ($LLR$) and is represented by; $p_{us}^{llr}$ leads to the dissemination of RERRs in AODV. On the other hand, DSR piggy-backs RERR messages along with next RREQs in the case of route re-discovery process, while these RERR messages are generated in the case of success of $PS$.

In AODV, after unsuccessful $RD$ and after detecting link breakage in DYMO, RERR messages are broadcasted by the node which detects any link break and route rediscovery process is started through source node.

\begin{eqnarray}
CE_{RM}^{AODV} = & CE_{HELLO} + |sgn\ \  lb_{RN}| \sum_{i=1}^{N_{llr}}i  \nonumber \\
                 & + |sgn\ \  p_{us}^{llr}| + \sum_{N_{rerr}}i
\end{eqnarray}

We also compare the performance of AODV without HELLO messages and refer it as AODV-LL, and in this case link layer feed back is used. $HELLO\_INTERVAL$ for AODV is $1s$, and $ALLOWED\_HELLO\_LOSS$ is $2$, thus a link can be considered as broken after expiration of $ALLOWED\_HELLO\_LOSS$ value. To increase the efficiency of AODV, quick detection of link breakage is needed. The energy cost for AODV-LL is given as:

\begin{eqnarray}
CE_{RM}^{AODV-LL} = & |sgn\ \  lb_{AR}| \sum_{i=1}^{N_{llr}}i  \nonumber \\
                    & + |sgn\ \  p_{us}^{llr}| + \sum_{N_{rerr}}i
\end{eqnarray}

Whereas, $CE_{HELLO}$ is energy cost of HELLO messages for link monitoring. Further details for this cost is available in \cite{9}. When link breakages of RN ($lb_{RN}$) occurs that cause initialization of $LLR$. After unsuccessful $LLR$, RERR messages are broadcasted in AODV, and $N_{rerr}$ represents the number of nodes that receives RERR messages.

\begin{equation}
CE_{HELLO} = \frac {\tau_{route\_in\_use}}{\tau_{H\_interval}} \times N_{RN}
\end{equation}

Let $\tau_{route\_in\_use}$ is the total time in which route remains in use, while $\tau_{H_interval}$ specifies the $HELLO\_INTERVAL$ (which is 1s in AODV). Moreover, $N_{RN}$ represent the number of nodes in Active Routes ($RN(s)$).

Like AODV, in case of DSR's $PS$ technique can reduce both the energy and time cost to be paid by a reactive protocol by diminishing the route re-discovery. In the case of successful $PS$, RERR messages are broadcasted to neighbors for the deletion of useless routes. Whereas, the absence of alternate route(s) in $RC$ leads to the failure of $PS$. In this situation, RERR messages are to be sent by piggy-backing them in the next RREQ messages during RD process.

\begin{equation}
CE_{RM}^{DSR}=  \sum_{i=1}^{n_{ps}} i
\end{equation}
where, ${n_{ps}}$ denote the node that salvage packet successfully.

\begin{equation}
CE_{RM}^{DYMO}= CE_{HELLO}+ |sgn\ \  lb_{AR}| \sum_{i=1}{N_{rerr}}i
\end{equation}

\section{Simulation Result}

We evaluate performance of the proposed framework in NS-2. For simulation setup, we choose Random Way point mobility model. We take mobilities and traffic flows scenarios for our evaluation. The area specified is $1000m \times 1000m$ field presenting a square space to allow mobile nodes to move inside. All of the nodes are provided with wireless links of a bandwidth of $2Mbps$ to transmit on. Simulations are run for $900s$ each. For evaluating mobilities effects, we vary pause time from $0s$ to $900s$ for $50$ nodes with in two different speeds of $2m/s$ and $30m/s$ separately. For evaluating different network flows with $15m/s$ speed and fixed pause time of $2s$, 1) different scalabilities from $10$ to $100$ nodes 2) traffic rate of $2$, $4$, $8$. $16$ and $32$ $packs/s$ for $50$ nodes.. We evaluate and compare the protocols by three performance parameters; throughput, $CT$ in terms of average end-to-end delay, and $CE$ in terms of normalized routing load.

\subsection{Throughput}

\begin{figure}[h]
\begin{center}
\includegraphics[width=70mm]{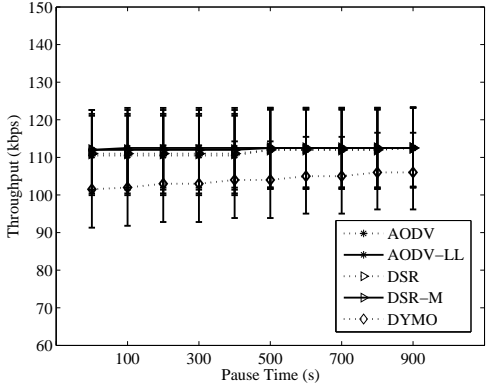}
\end{center}
\vspace{-0.6cm}
\caption{Throughput of Reactive Protocols at 2m/sec Mobility}
\label{fig:04}
\end{figure}

\begin{figure}[h]
\begin{center}
\includegraphics[width=70mm]{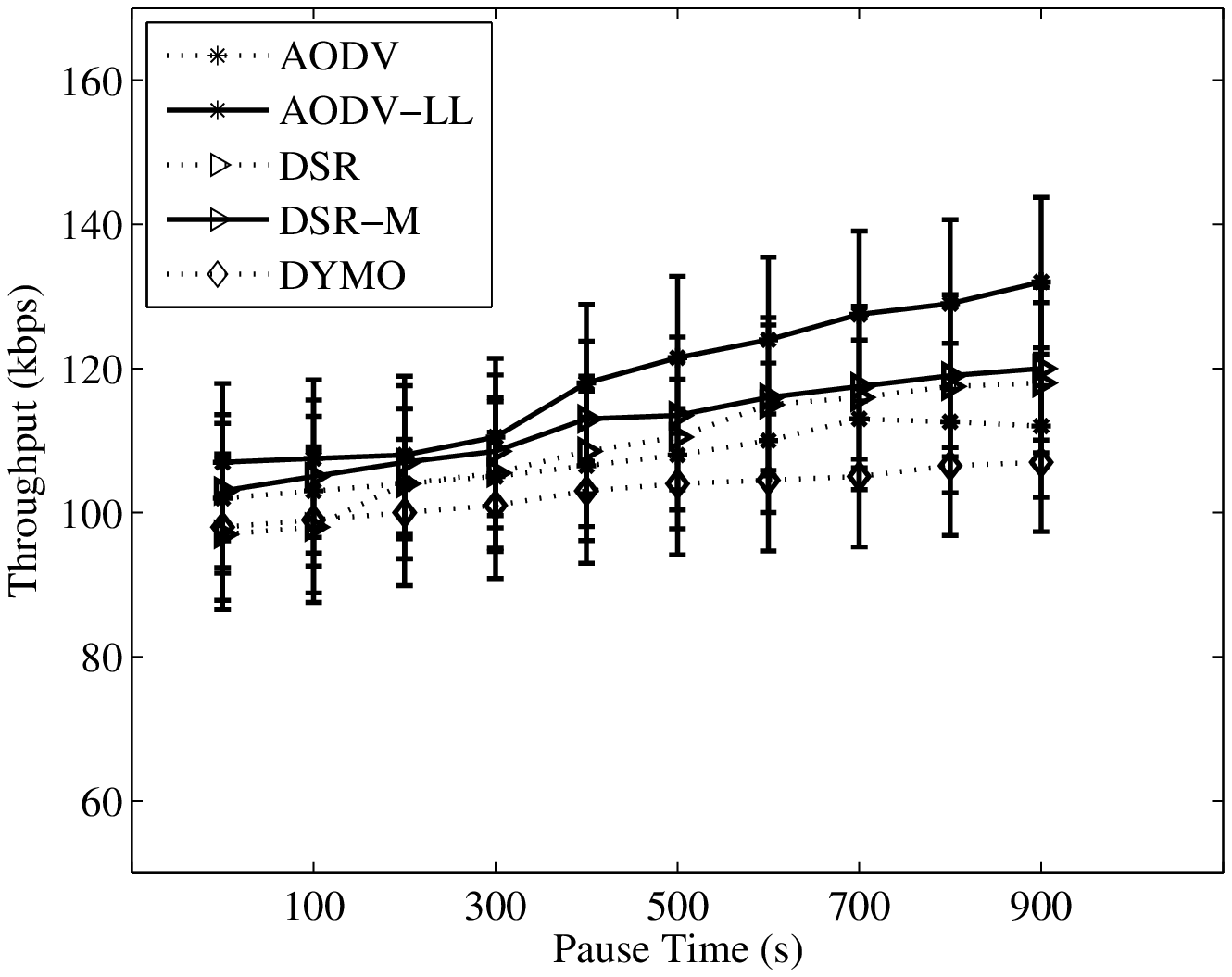}
\end{center}
\vspace{-0.6cm}
\caption{Throughput of Reactive Protocols at 30m/sec Mobility}
\label{fig:05}
\end{figure}

DSR gives high throughput in Fig. \ref{fig:04} because of accurate and efficient mechanisms for RD and RM processes due to low speeds of $2m/s$. From Fig. \ref{fig:05}, it is obvious that in very high dynamic situation, RC of DSR becomes ineffective, as, there is no mechanism to delete the stale routes from RC, and RERR messages are disseminated not traditionally as in other protocols; thus, the protocol fails to converge at this mobility speed. While AODV checks the route with valid time and avoids using the invalid routes from RT, thus, achieves more successful probability of RD.

In AODV-LL, quick detection and retirement make this protocol more efficient then AODV, and DSR-M reduces generation of stale routes information by reducing $TAPE\_CACHE\_SIZE$ (In DSR-M, we change $TAP\_CACHE\_SIZE$ from $1024$ to $256$, this modification results quick updation of RC).
Moreover, the HELLO messages and $LLR$ make able the protocol to handle highest rate of mobility, thus, overall converges in dynamic situations.  The worst behavior of DYMO among reactive protocols in response to mobility by showing overall less throughput value as is noticed in Fig. \ref{fig:04} and \ref{fig:05}. The absence of any supplementary mechanism make its performance low against respective constraints of throughput in high mobilities.

\begin{figure}[h]
\begin{center}
\includegraphics[width=70mm]{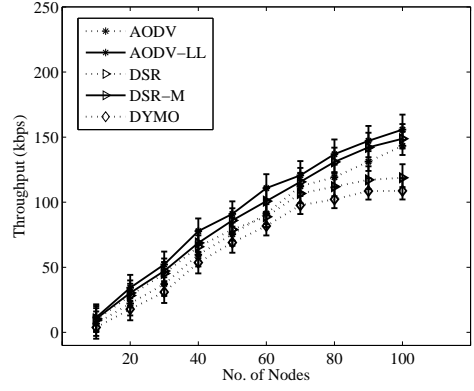}
\end{center}
\vspace{-0.6cm}
\caption{Throughput of Reactive Protocols vs Scalability}
\label{fig:06}
\end{figure}

\begin{figure}[h]
\begin{center}
\includegraphics[width=70mm]{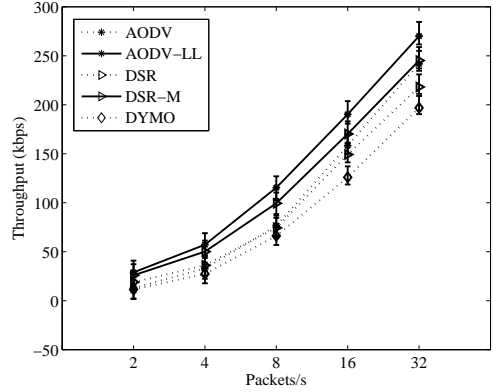}
\end{center}
\vspace{-0.6cm}
\caption{Throughput of Reactive Protocols vs Traffic Rate}
\label{fig:07}
\end{figure}

Conducted simulation results from Fig. \ref{fig:06} and \ref{fig:07}, AODV shows convergence for all data rates and all scalabilities, whereas DSR is less scalable while DYMO degrades its performance in more population of nodes. In \cite{4}, it is specified that `` AODV can better handle a wireless network of tens to thousand nodes '', therefore,  it performs better among reactive protocols for high network flows. The presence of \textit{grat. RREPs} and time-based routing activities that makes able the protocol to perform well by always choosing a fresher end-to-end path. The route deletion using RERR messages is also traditional and disseminates quick information after failure of $LLR$. It also maintains predecessor list; RERR packets reach all nodes using a failed link on its route to any desired destination.

\subsection{Cost of Time}

\begin{figure}[h]
\begin{center}
\includegraphics[width=70mm]{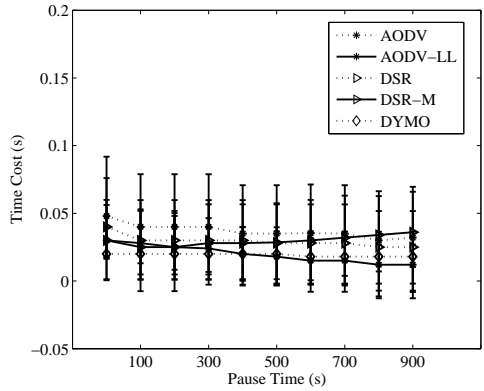}
\end{center}
\vspace{-0.6cm}
\caption{Time cost analysis of Reactive Protocols at 2m/sec Mobility}
\label{fig:08}
\end{figure}

\begin{figure}[h]
\begin{center}
\includegraphics[width=70mm]{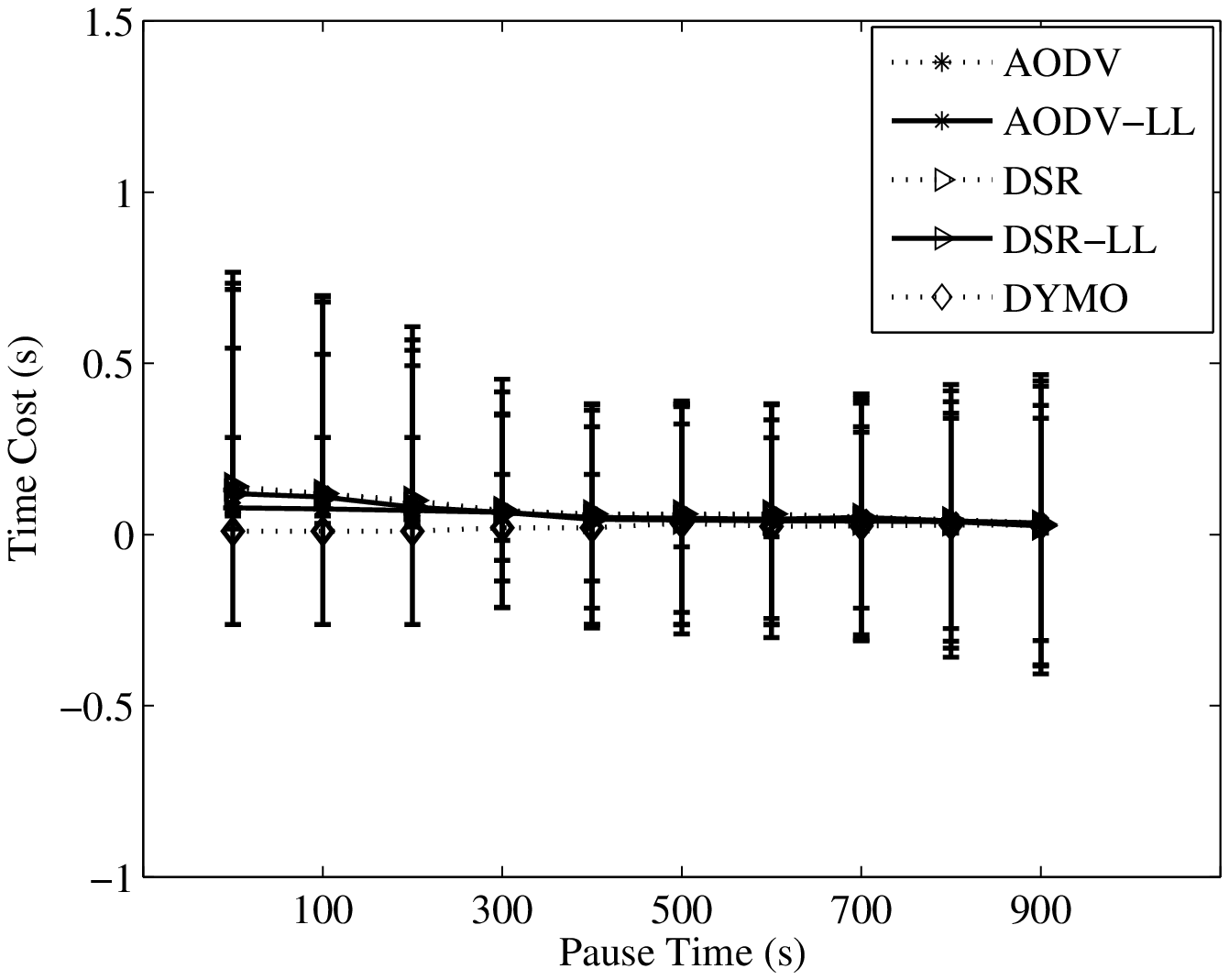}
\end{center}
\vspace{-0.6cm}
\caption{Time cost analysis of Reactive Protocols at 30m/sec Mobility}
\label{fig:09}
\end{figure}

\begin{figure}[h]
\begin{center}
\includegraphics[width=70mm]{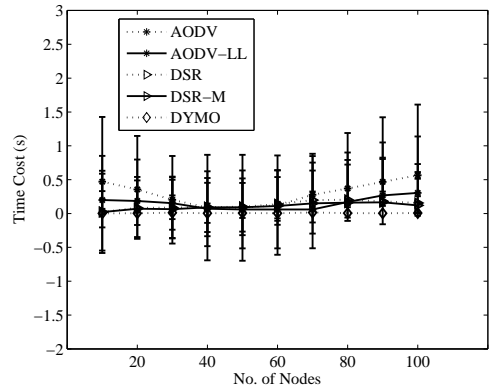}
\end{center}
\vspace{-0.6cm}
\caption{Time cost analysis of Reactive Protocols vs Scalability}
\label{fig:10}
\end{figure}

\begin{figure}[h]
\begin{center}
\includegraphics[width=70mm]{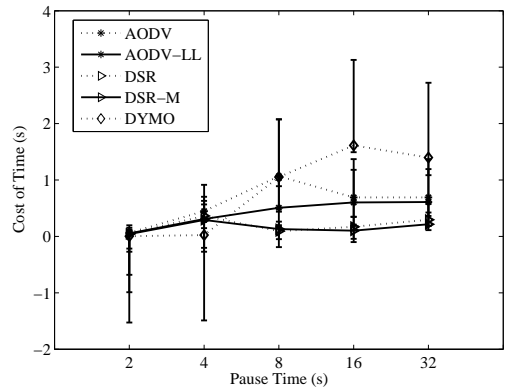}
\end{center}
\vspace{-0.6cm}
\caption{Time cost analysis of Reactive Protocols vs Traffic Rate}
\label{fig:11}
\end{figure}

AODV among reactive protocols in attains the highest delay. Because in local repair for link breaks in routes sometimes results in increased path lengths. In AODV-LL, E2ED becomes much less as compared to routing latency of AODV, because LLR initiation and repairement starts quickly after receiving link layer feed-back (link layer beacon messages to check the connectivity is send $100$ times in a second, and after $8$ connective failure notify link breakage), as depicted in Figs. \ref{fig:08},\ref{fig:09},\ref{fig:10} and \ref{fig:11}. DSR does not implement $LLR$ \cite{10},\cite{11}, therefore, its $CT$ value is less than AODV but during moderate and high mobility RC search fails frequently and results high routing delay.

At higher mobility, DSR suffers the higher $CT$ value, as portrayed in Fig. \ref{fig:09}. The reasons include: for RD, it first searches the desired route in the RC and then starts RD if the search fails, moreover, this searching is also performed during RM for PS process. Therefore, in high mobilities with high speeds, it does not give feasible solution, as shown in Fig. \ref{fig:09}. DYMO produces the lowest $CT$ value among reactive protocols because it only uses ERS for route finding which results low delay; as checking the RC (in DSR) and RT (in AODV) before RD cause delay of node traversal information. DSR-M also gives low value for routing latencies while considering high mobilities and scalabilities (in Figs. \ref{fig:09},\ref{fig:10} and \ref{fig:11}).

DYMO does not use any supplementary strategies like \textit{grat. RREPs}, PSing, RCing or LLR, therefore it suffers lowest delay in low traffic while produce high latency in high data rates. On the other hand, absence of the mechanism keeps the lowest delay cost of DYMO in all scalabilities, as shown in Fig. \ref{fig:09}. PS and \textit{grat. RREPs} keep the delay low in medium and high traffic scenarios for DSR but first checking the RC instead of simple ERS based RD process augments the delay when population increases, thus, more delay of DSR is presented in Fig. \ref{fig:09}, as compared to DYMO. AODV experiences the highest E2ED in all scalabilities due to LLR process (Fig. \ref{fig:09}).

\section{Conclusion}
This paper contributes LP\_models for WMhNs. To practically examine the respective constraints over reactive routing protocols, we select AODV, DSR and DYMO. We relate effects of RD and RM strategies of the selected protocols over WMhNs' constraints to check energy efficiency and delay reduction of chosen protocols in different scenarios in NS-2 while considering throughput, and cost of time. Quick route repair and optimizations of retransmission attempts result in better performance of the protocol by reducing energy utilization and routing latencies. For quick deletion of stale route entries in DSR, we reduce $TAP\_CACHE\_SIZE$ of DSR (DSR-M) and compare it with original DSR. For quick repairement, we compare AODV with and without link level feed back. Finally we deduce that AODV-LL due to quick repairment produces highest throughput by providing feasible solution for $max\,\,\,T_{avg}$ and  $min\,\,\,CT$.

\end{document}